\documentstyle[aps,prd,preprint,tighten,
mathrsfs,amsmath,amsfonts,epic,epsfig]{revtex}

\begin{document}

%\draft

\title{Three flavor neutrino oscillations in matter}

\author{Tommy Ohlsson\footnote{Electronic address: {\tt
tommy@theophys.kth.se}} and H{\aa}kan Snellman\footnote{Electronic
address: {\tt snell@theophys.kth.se}}}
\address{Division of Mathematical Physics, Theoretical Physics, Department of
Physics,\\ Royal Institute of Technology, SE-100 44 Stockholm, Sweden}
\date{Received 5 November 1999}

\maketitle

\begin{abstract}
We derive analytic expressions for three flavor neutrino oscillations
in the presence of matter in the plane wave approximation using the
Cayley--Hamilton formalism. Especially, we calculate the time evolution
operator in both flavor and mass bases. Furthermore, we find the
transition probabilities, matter mass squared differences, and matter
mixing angles all expressed in terms of the vacuum mass squared differences,
the vacuum mixing angles, and the matter density. The conditions for
resonance in the presence of matter are also studied in some examples.
\end{abstract}

%\pacs{PACS number(s): 14.60.Pq, 14.60.Lm, 13.15.+g, 96.40.Tv}

\section{Introduction}
\label{sec:intro}

There are at present essentially four different types of experiments
looking for neutrino oscillations \cite{pont57}: solar neutrino experiments,
atmospheric neutrino experiments, accelerator neutrino experiments,
and reactor neutrino experiments. Among the accelerator and reactor
experiments, the accelerator long baseline experiments \cite{LBL} are just
beginning to become operative. Also the Sudbury Neutrino Observatory
(SNO) \cite{SNO} has started to take data on solar neutrinos. An
accumulating amount of data on neutrino oscillations is therefore
becoming accessible. Yet, at present there is no general agreement on
the mixing angles or the mass squared differences of the neutrinos.

In a previous paper \cite{ohls99}, we described a global analysis of
neutrino oscillation data in a three flavor neutrino scenario.  In
that analysis, we deliberately considered mass ranges for the solution
that avoided the region where the solar and atmospheric neutrinos
could be affected by the so-called Mikheyev--Smirnov--Wolfenstein
(MSW) effect \cite{mikh85}.

Here we would like to report on the second stage of a global analysis
of all the neutrino oscillation data in the three flavor neutrino model and
present analytic expressions for the oscillation probabilities in the
presence of matter expressed in the Cabibbo--Kobayashi--Maskawa (CKM)
mixing matrix elements (or [vacuum] mixing angles) and the neutrino
energies or masses, i.e., incorporating the MSW effect.

We will assume as before that $CP$ nonconservation is negligible at the
present level of experimental accuracy \cite{dick99}. Thus, the CKM
matrix for the neutrinos is real.

Previous work on models for three flavor neutrino oscillations in
matter includes works by Barger {\it et al.} \cite{barg80}, Kim and
Sze \cite{kim87}, and Zaglauer and Schwarzer
\cite{zagl88}. Our method is different from all these approaches and their
parametrizations also differ slightly from ours. In particular, we
calculate the time evolution operator and do not introduce the
auxiliary (effective) matter mixing angles.

Approximative solutions for three flavor neutrino
oscillations in matter have been presented by Kuo and Pantaleone
\cite{kuo86} and Joshipura and Murthy \cite{josh88}. Approximative
treatments have also been done by Toshev and Petcov \cite{tosh87}.
D'Olivo and Oteo have made contributions by using an approximative
Magnus expansion for the time evolution operator \cite{oliv96} and
Aquino, Bellandi, and Guzzo have found the auxiliary matter mixing angles in
terms of the vacuum mixing angles \cite{aqui95}.
Extensive numerical investigations of matter enhanced three flavor
oscillations have been made by Fogli {\it et al.} \cite{fogl94}.

Below we derive the expressions for the transition probabilities in the case of
three flavor neutrino oscillations when the neutrinos pass through matter
(the MSW effect) using Cayley--Hamilton's theorem. This allows us to give
rather explicit analytic formulas in terms of elementary functions for
the quite involved expressions for the transition probabilities. We
derive the time evolution operator for a constant matter (electron)
density. The most straightforward way to use this result is to approximate any
density profile with step functions and to obtain the total time evolution
as a product of the evolution in each step. This can easily be handled
numerically \cite{freu99}.

The three auxiliary angles of rotation from the matter mass eigenstate
basis to the flavor state basis can be identified from these
expressions and we show their relation to the ordinary formalism.

The application of this formalism to the analysis of the neutrino
oscillation data will be presented in a forthcoming publication.
Here we will only illustrate the qualitative behavior of the
transition probabilities and the auxiliary matter mixing angles
as functions of the matter density for different choices of the vacuum
mixing angles and the mass squared differences.

It is possible to adapt the formalism presented here to the
case of a scenario with an arbitrary number of neutrinos, but the
eigenvalues have to be solved for numerically for the case of five or
more neutrino flavors.

The outline of our paper is as follows.  We first go through some
formalism in Sec.~\ref{sec:form} and then set out to find the solution
for the time evolution operator and the transition probabilities in
the presence of matter. In Sec.~\ref{sec:evol}, the main result of our analysis
is given by the time evolution operator for the neutrinos when passing
through matter with constant matter density expressed as a finite sum
of elementary functions in the matrix elements of the Hamiltonian, and
in Sec.~\ref{sec:prob}, the transition probabilities in presence of
matter are given. In Sec.~\ref{sec:mma}, we determine the auxiliary
matter mixing angles. Finally, in Sec.~\ref{sec:disc}, we present a
discussion of our main results and the illustrations.

\section{Formalism}
\label{sec:form}

Neutrinos are produced in flavor ``eigenstates'' $\vert \nu_\alpha
\rangle$ ($\alpha=e,\mu,\tau$) created by the interaction
of the weak gauge bosons with the charged leptons. Between the source,
the production point of the neutrinos, and the detector, the neutrinos
evolve as mass eigenstates $\vert \nu_a \rangle$ ($a=1,2,3$),
i.e., as states with definite mass.
At the detector, the neutrino flavors can again be identified by
charge current interactions.  It is also possible to detect the
neutrinos by weak neutral current interactions.  In this case,
however, the weak neutral currents do not distinguish the flavors on
the detector side, and the total neutrino flux will be measured.

Let the flavor state basis and the mass eigenstate basis be given by
${\cal H}_{f} \equiv \{ \vert \nu_\alpha \rangle
\}_{\alpha=e,\mu,\tau}$ and ${\cal H}_{m} \equiv \{ \vert \nu_a \rangle
\}_{a=1}^3$, respectively. Then the flavor
states $\vert \nu_\alpha \rangle \in {\cal H}_{f}$ can be  obtained as
superpositions of the mass eigenstates $\vert \nu_a \rangle \in {\cal
H}_{m}$, or vice versa. Observe that the bases ${\cal H}_f$ and ${\cal
H}_m$ are just two different representations of the same Hilbert space
${\cal H}$.

In the present analysis, we will use the plane wave approximation to
describe neutrino oscillations. In this approximation, a neutrino
field $\nu_{\alpha}$ with flavor $\alpha$ is a linear combination of
neutrino mass fields $\nu_{a}$ such that \cite{kim93}
\begin{equation}
\nu_{\alpha} = \sum_{a=1}^3 U_{\alpha a} \nu_{a}, \quad \alpha = e,
\mu, \tau,
\label{eq:fields}
\end{equation}
where the $U_{\alpha a}$'s are entries in a $3 \times 3$ unitary
matrix $U$. Taking the Hermitian conjugate of Eq.~(\ref{eq:fields})
yields
\begin{equation}
\nu^\dagger_{\alpha} = \sum_{a=1}^3 U^\ast_{\alpha a} \nu^\dagger_{a},
\quad \alpha = e, \mu, \tau.
\label{eq:fields_herm}
\end{equation}
Acting with Eq.~(\ref{eq:fields_herm}) on the vacuum state $\vert 0
\rangle$ gives
\begin{equation}
\vert \nu_\alpha \rangle = \sum_{a=1}^3 U^\ast_{\alpha a} \vert \nu_a
\rangle, \quad \alpha = e, \mu, \tau,
\end{equation}
since the states are defined as $\vert \nu_\alpha \rangle \equiv
\nu^\dagger_\alpha \vert 0 \rangle$, where $\alpha = e,\mu,\tau$, and $\vert
\nu_a \rangle \equiv \nu^\dagger_a \vert 0 \rangle$, where $a =
1,2,3$. In what follows, we will use the short-hand notations $\vert
\alpha \rangle \equiv \vert \nu_\alpha \rangle$ and $\vert a \rangle
\equiv \vert \nu_a \rangle$ for the flavor states and the mass
eigenstates, respectively.
An arbitrary neutrino state $\psi \in {\cal H}$ can, of course, be
expressed in both the flavor and mass bases as
\begin{equation}
\psi \equiv \sum_{\alpha=e,\mu,\tau} \psi_\alpha \vert \alpha \rangle =
\sum_{\alpha=e,\mu,\tau} \psi_\alpha \sum_{a=1}^3 U^\ast_{\alpha a}
\vert a \rangle = \sum_{a=1}^3 \left( \sum_{\alpha=e,\mu,\tau}
\psi_\alpha U^\ast_{\alpha a} \right) \vert a \rangle \equiv
\sum_{a=1}^3 \psi_a \vert a \rangle,
\end{equation}
where $\psi_\alpha$, $\alpha=e,\mu,\tau$, and $\psi_a$, $a=1,2,3$, are
the components of the state $\psi$ in the flavor state basis and the mass
eigenstate basis, respectively, and they are related to each other by
\begin{equation}
\psi_a = \sum_{\alpha=e,\mu,\tau} U^\ast_{\alpha a} \psi_\alpha, \quad a=1,2,3.
\end{equation}

In matrix form, one has
\begin{equation}
\psi_f = U \psi_m,
\end{equation}
where
$$
\psi_f = \left( \psi_\alpha \right) = \left( \begin{array}{c} \psi_e
\\ \psi_\mu \\ \psi_\tau
\end{array} \right) \in {\cal H}_f \quad \mbox{and} \quad \psi_m =
\left( \psi_a \right) = \left(
\begin{array}{c} \psi_1 \\ \psi_2 \\ \psi_3 \end{array} \right) \in
{\cal H}_m.
$$
Here the subscripts $f$ and $m$ denote the flavor state basis and the
mass eigenstate basis, respectively.

The unitary matrix $U$ is given by
\begin{equation}
U = (U_{\alpha a}) = \left( \begin{array}{ccc} U_{e 1} & U_{e 2} &
U_{e 3} \\ U_{\mu 1} & U_{\mu 2} & U_{\mu 3} \\ U_{\tau 1} & U_{\tau
2} & U_{\tau 3} \end{array} \right).
\end{equation}
A convenient parametrization for $U = U(\theta_1,\theta_2,\theta_3)$
is given by \cite{caso98}
\begin{equation}
U = \left( \begin{array}{ccc} C_2 C_3 & S_3 C_2 & S_2 \\ - S_3 C_1 -
S_1 S_2 C_3 & C_1 C_3 - S_1 S_2 S_3 & S_1 C_2 \\ S_1 S_3 - S_2
C_1 C_3 & - S_1 C_3 - S_2 S_3 C_1 & C_1 C_2 \end{array} \right),
\end{equation}
where $S_i \equiv \sin \theta_i$ and $C_i \equiv \cos \theta_i$ for $i
= 1,2,3$. This is the so-called standard representation of the
CKM mixing matrix. The quantities $\theta_i$, where $i = 1,2,3$, are
the so-called (vacuum) mixing angles. We have here put the $CP$
phase equal to zero in the CKM matrix. This means that $U^\ast_{\alpha
a} = U_{\alpha a}$ for $\alpha = e, \mu, \tau$ and $a = 1,2,3$.

In the mass eigenstate basis ${\cal H}_m$, the Hamiltonian ${\mathscr
H}$ for the propagation of the neutrinos in vacuum is diagonal and given by
\begin{equation}
H_m = \left( \begin{array}{ccc} E_{1} & 0 & 0 \\ 0 & E_{2} & 0 \\ 0 &
 0 & E_{3} \end{array} \right),
\end{equation}
where $E_a = \sqrt{m_a^2 + {\bf p}^2}$, $a = 1,2,3$, are the energies
of the neutrino mass eigenstates $\vert a \rangle$, $a = 1,2,3$ with
masses $m_a$, $a = 1,2,3$. Note that we assume the momentum ${\bf p}$ to be the
same for all mass eigenstates.

When neutrinos propagate in matter, there is an additional term coming
from the presence of electrons in the matter \cite{mikh85}. This term
is diagonal in the flavor state basis ${\cal H}_f$ and is given by
\begin{equation}
V_{f} = A \left( \begin{array}{ccc} 1 &0 & 0 \\0 & 0 & 0 \\ 0 & 0 &
 0\end{array} \right),
\end{equation}
where $A = \pm \sqrt{2} G_{\rm F} N_e$ is the matter density. Here
$G_{\rm F}$ is the Fermi weak coupling constant and $N_e$ is the electron
density. The sign depends on whether we deal with neutrinos~($+$) or
antineutrinos~($-$). We will assume that the electron density $N_e$ is
constant throughout the matter in which the neutrinos are propagating.
In the mass representation, this piece of the
Hamiltonian is $V_{m}=U^{-1}V_{f}U$, where again $U$ is the CKM matrix.

We are looking for the unitary transformation that leads from the
initial state $\psi_{f}(0)$ in flavor basis at time $t=0$ of
production of the neutrino, to the state of the same neutrino
$\psi_{f}(t)$ at the detector at time $t$.  This transformation is
given by the operator $U_{f}(t)\equiv U_{f}(t,0)$, where
$U_{f}(t_{2},t_{1})$ is the time evolution operator from time $t_{1}$ to time
$t_{2}$ in flavor basis.  This operator can be formally written as
$U_{f}(t)=e^{-iH_{f}t}$.  When the neutrinos are propagating through
vacuum, the Hamiltonian in flavor basis is $H_{f}= UH_{m}U^{-1}$.  In
this case, the exponentiation of $H_{f}$ can be effectuated
easily: $U_{f}(t)=e^{-iH_f t} = Ue^{-iH_{m}t}U^{-1}$, and the result
can be expressed in closed form.  In the case when the neutrinos
propagate through matter, the Hamiltonian is not diagonal in either
the mass eigenstate or the flavor state basis, and we have to calculate the
operator $U_{f}(t)$.  This is the task we now set out for.

\section{Calculating the time evolution operator}
\label{sec:evol}

The Schr{\"o}dinger equation in the mass eigenstate basis is
\begin{equation}
i \frac{d}{dt} \psi_m(t) = {\mathscr H}_m \psi_m(t),
\label{eq:schr_m}
\end{equation}
where
$$
{\mathscr H}_m = H_m + V_m
$$
is the Hamiltonian.

Similarly, in the flavor state basis, one has
\begin{equation}
i \frac{d}{dt} \psi_f(t) = {\mathscr H}_f \psi_f(t),
\end{equation}
where
$$
{\mathscr H}_f = H_f + V_f.
$$
In the mass eigenstate basis, the matter term (the potential matrix)
is
\begin{equation}
V_m = U^{-1} V_f U.
\label{eq:V_m}
\end{equation}
Thus, the total Hamiltonian in the mass eigenstate basis is given by
\begin{equation}
{\mathscr H}_m = H_m + U^{-1} V_f U.
\end{equation}
Equation~(\ref{eq:schr_m}) has the solution
\begin{equation}
\psi_m(t) = e^{-i {\mathscr H}_m t} \psi_m(0).
\label{eq:fin_psi_m(t)}
\end{equation}
Inserting $t = L$ into Eq.~(\ref{eq:fin_psi_m(t)}), one finds the
solution in the mass eigenstate basis to be
\begin{equation}
\psi_m(L) = e^{-i {\mathscr H}_m L} \psi_m(0) \equiv U_m(L) \psi_m(0)
\end{equation}
and in the flavor state basis
\begin{eqnarray}
\psi_f(L) &=& U \psi_m(L) = U e^{-i {\mathscr H}_m L}
\psi_m(0) = U e^{-i {\mathscr H}_m L} U^{-1} U
\psi_m(0) = U e^{-i {\mathscr H}_m L} U^{-1} \psi_f(0) \nonumber\\
&\equiv& U_f(L) \psi_f(0).
\end{eqnarray}

Inserting the potential matrix in the flavor basis and the CKM matrix
into Eq.~(\ref{eq:V_m}) yields
\begin{equation}
V_m = A \left( \begin{array}{ccc} U_{e1}^2 & U_{e1} U_{e2} & U_{e1}
U_{e3} \\ U_{e2} U_{e1} & U_{e2}^2 & U_{e2} U_{e3} \\ U_{e3} U_{e1} &
U_{e3} U_{e2} & U_{e3}^2 \end{array} \right),
\end{equation}
which is a (real) symmetric matrix. Then, the Hamiltonian in the mass
eigenstate basis is
\begin{equation}
{\mathscr H}_m = \left( \begin{array}{ccc} E_1 + A U_{e1}^2 & A U_{e1}
U_{e2} & A U_{e1} U_{e3} \\ A U_{e2} U_{e1} & E_2 + A U_{e2}^2 & A
U_{e2} U_{e3} \\ A U_{e3} U_{e1} & A U_{e3} U_{e2} & E_3 + A U_{e3}^2
\end{array} \right).
\end{equation}
Of course, we also have $U_{f}(L) = e^{-i{\mathscr H}_{f}L}$, but we
prefer to first work out $U_{m}(L) = e^{-i{\mathscr H}_{m}L}$, since
the structure of ${\mathscr H}_m$ is much simpler than that of
${\mathscr H}_f$, and afterwards perform the transformation to $U_{f}(L)$.

Next, we need to find the exponential of the matrix
$-i {\mathscr H}_m L$ in order to obtain $U_m(L)$. Generally, when one
wants to find the exponential of an $N \times N$ matrix $M$, one has
to calculate
\begin{equation}
e^M = \sum_{n = 0}^{\infty} \frac{1}{n!} M^n.
\label{eq:inf_series}
\end{equation}
The characteristic equation of the matrix $M$ is given by
\begin{equation}
\chi(\lambda) \equiv \det (M - \lambda I) = \lambda^N + c_{N-1} \lambda^{N-1}
+ \cdots + c_1 \lambda + c_0 = 0,
\end{equation}
where $c_i$, $i = 0, 1, \ldots, N-1$, are some coefficients.
Cayley--Hamilton's theorem implies that $\lambda$ in the
characteristic equation can be replaced with the matrix $M$ itself,
giving
\begin{equation}
M^N + c_{N-1} M^{N-1} + \cdots + c_1 M + c_0 I = 0,
\end{equation}
where $I$ is the $N \times N$ identity matrix. Hence, one has
\begin{equation}
M^N = -c_{N-1} M^{N-1} - \cdots - c_1 M - c_0 I,
\end{equation}
which means that
\begin{equation}
M^p = c_{N-1}^{(p)} M^{N-1} + \cdots + c_1^{(p)} M + c_0^{(p)} I,
\quad \mbox{for any $p \geq N$},
\end{equation}
where $c_i^{(p)}$, $i = 0,1,\ldots,N-1$, are some coefficients.
The exponential of the matrix $M$ can then be written as
\begin{equation}
e^M = a_0 I + a_1 M + \cdots + a_{N-1} M^{N-1} = \sum_{n=0}^{N-1} a_n M^n,
\label{eq:eM}
\end{equation}
where $a_i$, $i = 0,1,\ldots,N-1$, are some coefficients. Thus, the
infinite series in Eq.~(\ref{eq:inf_series}) is reduced to only $N$
terms, where $N$ is the dimension of the matrix $M$.

An arbitrary $N \times N$ matrix $M$ can always be written
as
\begin{equation}
M = M_0 + \frac{1}{N} ({\rm tr \,} M) I,
\label{eq:traceless}
\end{equation}
where $M_0$ is an $N \times N$ traceless matrix, i.e., ${\rm tr
\,}M_0 = 0$. Splitting up $M$ as in Eq.~(\ref{eq:traceless}) is useful, since
the identity matrix $I$ is commuting with all other matrices.

Now using Eqs.~(\ref{eq:eM}) and (\ref{eq:traceless}), one can write
the exponential of the matrix $-i {\mathscr H}_m L$ as
\begin{equation}
e^{-i {\mathscr H}_m L} = \phi e^{-i LT} =
\phi \left[ a_0 I + a_1 (-i LT) + a_2 (-i LT)^2 \right] =
\phi \left( a_0 I - i LT a_1 - L^2 T^2 a_2 \right),
\label{eq:fin_series}
\end{equation}
where $\phi \equiv e^{-i L {\rm tr \,}{\mathscr H}_m/3}$ is a complex
phase factor, $T \equiv {\mathscr H}_m - ({\rm tr \,}{\mathscr H}_m)
I/3$ is a traceless matrix, and $a_0$, $a_1$, and $a_2$ are coefficients to be
determined. The infinite series in Eq.~(\ref{eq:inf_series})
is here reduced to only three terms. The trace of the matrix
${\mathscr H}_m$ is
\begin{equation}
{\rm tr\,} {\mathscr H}_m = E_1 + E_2 + E_3 + A.
\end{equation}
The matrix $T$ can now be written as
\begin{eqnarray}
T &=& (T_{ab}) \nonumber\\
&=& \left( \begin{array}{ccc} A U_{e1}^2 - \tfrac{1}{3} A +
\tfrac{1}{3} \left( E_{12} + E_{13} \right) & A U_{e1} U_{e2} & A
U_{e1} U_{e3} \\ A U_{e1} U_{e2} & A U_{e2}^2 - \tfrac{1}{3} A +
\tfrac{1}{3} \left( E_{21} + E_{23} \right) & A U_{e2} U_{e3} \\ A
U_{e1} U_{e3} & A U_{e2} U_{e3} & A U_{e3}^2 - \tfrac{1}{3} A +
\tfrac{1}{3} \left( E_{31} + E_{32} \right)
\end{array} \right), \nonumber\\
\end{eqnarray}
where $E_{ab} \equiv E_a - E_b$. The six quantities $E_{ab}$, where
$a,b=1,2,3$ and $a \neq b$, are not linearly independent, since they
obey the following relations:
\begin{equation}
E_{ba} = - E_{ab},
\end{equation}
\begin{equation}
E_{12} + E_{23} + E_{31} = 0.
\end{equation}
This means that only two of the $E_{ab}$'s are linearly independent,
e.g., $E_{21}$ and $E_{32}$.

The following linear system of equations will determine the
coefficients $a_0$, $a_1$, and $a_2$ introduced in
Eq.~(\ref{eq:fin_series}) above,
\begin{equation}
\begin{array}{l} e^{-i L \lambda_1} = a_0 - i L \lambda_1 a_1
- L^2 \lambda_1^2 a_2 \\
e^{-i L \lambda_2} = a_0 - i L \lambda_2 a_1 - L^2 \lambda_2^2 a_2 \\
e^{-i L \lambda_3} = a_0 - i L \lambda_3 a_1 - L^2 \lambda_3^2 a_2
\end{array},
\label{eq:eqsys}
\end{equation}
where $\lambda_a$, $a=1,2,3$, are the
eigenvalues of the matrix $T$, i.e., solutions to the
characteristic equation
$$
\lambda^3 + c_2 \lambda^2 + c_1 \lambda + c_0 = 0
$$
with
$$
c_2 = - {\rm tr \,}T = 0,
$$
\begin{eqnarray}
c_1 &=& T_{11} T_{22} - T_{12} T_{21} + T_{11} T_{33} - T_{13} T_{31}
+ T_{22} T_{33} - T_{23} T_{32} \nonumber\\
&=& - \tfrac{1}{3} A^2 + \tfrac{1}{3} A \left[ U_{e1}^2 \left( E_{21} +
E_{31} \right) + U_{e2}^2 \left( E_{12} + E_{32} \right) + U_{e3}^2
\left( E_{13} + E_{23} \right) \right] \nonumber\\
&-& \tfrac{1}{9} \left(
E_{12}^2 + E_{13}^2 + E_{23}^2 + E_{12} E_{13} + E_{21} E_{23} +
E_{31} E_{32} \right), \nonumber
\end{eqnarray}
\begin{eqnarray}
c_0 = - \det T &=& - \tfrac{2}{27} A^3 + \tfrac{1}{9} A^2 \left[ U_{e1}^2
\left( E_{21} + E_{31} \right) + U_{e2}^2 \left( E_{12} + E_{32}
\right) + U_{e3}^2 \left( E_{13} + E_{23} \right) \right] \nonumber\\
&-& \tfrac{1}{9} A \big[ U_{e1}^2 \left( E_{21} + E_{23} \right)
\left( E_{31} + E_{32} \right) + U_{e2}^2 \left( E_{12} + E_{13}
\right) \left( E_{31} + E_{32} \right) \nonumber\\
&+& U_{e3}^2 \left( E_{12} + E_{13} \right) \left( E_{21} + E_{23}
\right) + \tfrac{1}{3} \left( E_{12}^2 + E_{13}^2 + E_{23}^2 + E_{12} E_{13}
+ E_{21} E_{23} + E_{31} E_{32} \right) \big] \nonumber\\
&-& \tfrac{1}{27} \left( E_{12} + E_{13} \right) \left( E_{21} +
E_{23} \right) \left( E_{31} + E_{32} \right). \nonumber
\end{eqnarray}
Note that the coefficients $c_0$, $c_1$, and $c_2$ are all
real. Introducing the relativistic limit (see Appendix~\ref{app:rel}),
$E_{21}$ and $E_{32}$ can be written in terms of the mass squared
differences $\Delta m^2$ ($=\Delta m_{21}^2$) and $\Delta M^2$
($=\Delta m_{32}^2$) as
$$
E_{21} = \frac{\Delta m^2}{2E} \quad \mbox{and} \quad E_{32} \simeq
\frac{\Delta M^2}{2E},
$$
where $E$ is the neutrino energy.

The solutions to the characteristic equation are \cite{abra68}
\begin{equation}
\lambda_1 = - \frac{1}{2} \left( s_1 + s_2 \right) - \frac{1}{3} c_2
+ i \frac{\sqrt{3}}{2} \left( s_1 - s_2 \right) = - \frac{1}{2}
\left( s_1 + s_2 \right) + i \frac{\sqrt{3}}{2}
\left( s_1 - s_2 \right),
\end{equation}
\begin{equation}
\lambda_2 = - \frac{1}{2} \left( s_1 + s_2 \right) - \frac{1}{3} c_2
- i \frac{\sqrt{3}}{2} \left( s_1 - s_2 \right) = - \frac{1}{2}
\left( s_1 + s_2 \right) - i \frac{\sqrt{3}}{2} \left( s_1 - s_2
\right),
\end{equation}
\begin{equation}
\lambda_3 = s_1 + s_2 - \frac{1}{3} c_2 = s_1 + s_2,
\end{equation}
where
$$
s_1 = \left[ r + (q^3 + r^2)^{1/2} \right]^{1/3}, \quad s_2 = \left[ r -
(q^3 + r^2)^{1/2} \right]^{1/3},
$$
and
$$
q = \tfrac{1}{3} c_1 - \tfrac{1}{9} c_2^2 = \tfrac{1}{3} c_1,
$$
$$
r = \tfrac{1}{6} (c_1 c_2 - 3 c_0) - \tfrac{1}{27} c_2^3 = - \tfrac{1}{2} c_0.
$$
Since $T$ is a Hermitian matrix ($T$ is a real symmetric matrix), it
holds that $q^3 + r^2 \leq 0$ and all eigenvalues $\lambda_a$, where
$a=1,2,3$, are real. Furthermore, inserting $q = c_1/3$ and $r =
-c_0/2$ into the inequality $q^3 + r^2 \leq 0$, one finds $c_1^3/27 +
c_0^2/4 \leq 0$ from which one can conclude that $c_1 \leq 0$ and
$\vert c_1 \vert \geq 3 \vert c_0 \vert^{2/3}/\sqrt[3]{4}$, since
$c_0,c_1 \in {\mathbb R}$.
Thus, the roots (eigenvalues) can be written as
\begin{eqnarray}
\lambda_1 &=& - \sqrt{-\frac{1}{3} c_1} \cos \left[ \frac{1}{3}
\arctan \left( \frac{1}{c_0} \sqrt{-c_0^2 - \frac{4}{27} c_1^3}
\right) \right] + \sqrt{-c_1} \sin \left[ \frac{1}{3} \arctan \left(
\frac{1}{c_0} \sqrt{-c_0^2 - \frac{4}{27} c_1^3} \right) \right], \nonumber\\
\label{eq:l1}
\end{eqnarray}
\begin{eqnarray}
\lambda_2 &=& - \sqrt{-\frac{1}{3} c_1} \cos \left[ \frac{1}{3} \arctan
\left( \frac{1}{c_0} \sqrt{-c_0^2 - \frac{4}{27} c_1^3} \right)
\right] - \sqrt{-c_1} \sin \left[ \frac{1}{3} \arctan \left(
\frac{1}{c_0} \sqrt{-c_0^2 - \frac{4}{27} c_1^3} \right) \right], \nonumber\\
\label{eq:l2}
\end{eqnarray}
\begin{eqnarray}
\lambda_3 &=& 2 \sqrt{-\frac{1}{3} c_1} \cos \left[ \frac{1}{3} \arctan
\left( \frac{1}{c_0} \sqrt{-c_0^2 - \frac{4}{27} c_1^3} \right)
\right]. \label{eq:l3}
\end{eqnarray}
Observe that $\lambda_1 + \lambda_2 + \lambda_3 = 0$. This relation is
satisfied because of the fact that the trace of $T$ is
zero. In addition, the relations
$\lambda_1 \lambda_2 + \lambda_1 \lambda_3 + \lambda_2 \lambda_3 = c_1
\leq 0$ and $\lambda_1 \lambda_2 \lambda_3 = - c_0$ are fulfilled.

The $\lambda_a$'s are, up to a constant, the energy eigenvalues of the
neutrinos in the presence of matter. The measurable quantities are the energy
differences $\vert \lambda_a - \lambda_b \vert$, which in the
relativistic limit are related to the effective matter mass squared
differences $\Delta \tilde{m}_{ab}^2$ as
\begin{equation}
\vert \lambda_a - \lambda_b \vert = \frac{\vert \Delta
\tilde{m}_{ab}^2 \vert}{2 E}, \qquad a,b=1,2,3, \quad a \neq b,
\end{equation}
where $E$ is again the neutrino energy.

Introducing the following matrix form representation for the system of
equations in Eq.~(\ref{eq:eqsys}),
\begin{equation}
{\bf e} = \Lambda {\bf a},
\end{equation}
where
$$
{\bf e} = \left( \begin{array}{c} e^{-i L \lambda_1} \\ e^{-i L
\lambda_2} \\ e^{-i L \lambda_3} \end{array} \right), \quad
\Lambda = \left( \begin{array}{ccc} 1 & -i L \lambda_1 & - L^2
\lambda_1^2 \\ 1 & -i L \lambda_2 & - L^2 \lambda_2^2 \\ 1 & -i L
\lambda_3 & - L^2 \lambda_3^2 \end{array} \right), \quad \mbox{and}
\quad {\bf a} = \left( \begin{array}{c} a_0 \\ a_1 \\ a_2 \end{array} \right),
$$
one obtains the solution
\begin{eqnarray}
{\bf a} &=& \Lambda^{-1} {\bf e} \nonumber\\
&=& \frac{1}{D} \left( \begin{array}{ccc}
i L^3 \left( \lambda_2 \lambda_3^2 - \lambda_2^2 \lambda_3
\right) & i L^3 \left( \lambda_1^2 \lambda_3
- \lambda_1 \lambda_3^2 \right) & i L^3 \left( \lambda_1
\lambda_2^2 - \lambda_1^2 \lambda_2 \right) \\ - L^2 \left(
\lambda_2^2 - \lambda_3^2 \right) & - L^2 \left( \lambda_3^2 -
\lambda_1^2 \right) & - L^2 \left( \lambda_1^2 - \lambda_2^2 \right)
\\ -i L \left( \lambda_3 -
\lambda_2 \right) & -i L \left( \lambda_1 - \lambda_3 \right) & -i L
\left( \lambda_2 - \lambda_1 \right) \end{array}
\right) \left(
\begin{array}{c} e^{-i L \lambda_1} \\ e^{-i L \lambda_2} \\ e^{-i L
\lambda_3} \end{array} \right) \nonumber\\
&=& \frac{1}{D} \left( \begin{array}{c} i L^3 \left[ e^{-i L
\lambda_1} \left( \lambda_2 \lambda_3^2 - \lambda_2^2 \lambda_3
\right) + e^{-i L
\lambda_2} \left( \lambda_1^2 \lambda_3 - \lambda_1 \lambda_3^2
\right) + e^{-i L
\lambda_3} \left( \lambda_1 \lambda_2^2 - \lambda_1^2 \lambda_2
\right)\right] \\ - L^2 \left[ e^{-i L
\lambda_1} \left( \lambda_2^2 - \lambda_3^2
\right) + e^{-i L
\lambda_2} \left( \lambda_3^2 - \lambda_1^2
\right) + e^{-i L
\lambda_3} \left( \lambda_1^2 - \lambda_2^2
\right)\right] \\ -i L \left[ e^{-i L
\lambda_1} \left( \lambda_3 - \lambda_2
\right) + e^{-i L
\lambda_2} \left( \lambda_1 - \lambda_3
\right) + e^{-i L
\lambda_3} \left( \lambda_2 - \lambda_1
\right)\right] \end{array} \right), \nonumber\\
\end{eqnarray}
where
$$
D \equiv \det \Lambda = i L^3 \left(\lambda_1 \lambda_2^2 -
\lambda_1^2 \lambda_2 + \lambda_2 \lambda_3^2 - \lambda_2^2 \lambda_3
+ \lambda_3 \lambda_1^2 - \lambda_3^2 \lambda_1\right).
$$
Thus, the exponential of the matrix $-i {\mathscr H}_m L$ in
Eq.~(\ref{eq:fin_series}) can be written as
\begin{eqnarray}
e^{-i {\mathscr H}_m L} = \phi e^{-i L T} &=& -i \phi \frac{L^3}{D}
\bigg\{ \big[ e^{-i L \lambda_1}
\left( \lambda_2^2 \lambda_3 - \lambda_2 \lambda_3^2 \right) + e^{-i L
\lambda_2} \left( \lambda_1 \lambda_3^2 - \lambda_1^2 \lambda_3
\right) \nonumber\\
&+& e^{-i L \lambda_3} \left( \lambda_1^2 \lambda_2 - \lambda_1
\lambda_2^2 \right) \big] I + \big[ e^{-i L \lambda_1}
\left(\lambda_3^2- \lambda_2^2 \right) + e^{-i L \lambda_2} \left(
\lambda_1^2 - \lambda_3^2 \right) \nonumber\\
&+& e^{-i L \lambda_3} \left( \lambda_2^2 - \lambda_1^2 \right) \big] T
+ \big[ e^{-i L \lambda_1} \left( \lambda_3 - \lambda_2 \right) +
e^{-i L \lambda_2} \left( \lambda_3 - \lambda_1 \right) \nonumber\\
&+& e^{-i L \lambda_3} \left( \lambda_1 - \lambda_2 \right) \big] T^2 \bigg\}.
\label{eq:eiVmL}
\end{eqnarray}
Inserting the expression for $D$ into Eq.~(\ref{eq:eiVmL}) gives
\begin{eqnarray}
e^{-i {\mathscr H}_m L} &=& \frac{1}{(\lambda_1 -
\lambda_2)(\lambda_1 - \lambda_3)} \phi e^{-i L \lambda_1} \left[ \lambda_2
\lambda_3 I - (\lambda_2 + \lambda_3) T + T^2 \right] \nonumber \\
&+& \frac{1}{(\lambda_2 - \lambda_1)(\lambda_2 - \lambda_3)} \phi e^{-i L
\lambda_2} \left[ \lambda_1 \lambda_3 I - (\lambda_1 + \lambda_3) T +
T^2 \right] \nonumber \\
&+& \frac{1}{(\lambda_3 - \lambda_1)(\lambda_3 - \lambda_2)} \phi e^{-i L
\lambda_3} \left[ \lambda_1 \lambda_2 I - (\lambda_1 + \lambda_2) T +
T^2 \right].
\label{eq:eiH_mL}
\end{eqnarray}
Using the various relations for the eigenvalues finally yields
\begin{eqnarray}
e^{-i {\mathscr H}_m L} &=&
\phi e^{-i L \lambda_1} \frac{(\lambda_1^2
+ c_1) I + \lambda_1 T + T^2}{3\lambda_1^2+c_1}
+ \phi e^{-i L \lambda_2} \frac{(\lambda_2^2 + c_1) I + \lambda_2 T +
T^2}{3\lambda_2^2+c_1} \nonumber\\
&+& \phi e^{-i L \lambda_3} \frac{(\lambda_3^2 + c_1) I + \lambda_3 T
+ T^2}{3\lambda_3^2+c_1},
\label{eq:eiH_mL_fin}
\end{eqnarray}
which can be written as
\begin{equation}
U_m(L) = e^{-i {\mathscr H}_m L} = \phi \sum_{a=1}^3 e^{-i L \lambda_a}
\frac{1}{3\lambda_a^2+c_1} \left[ (\lambda_a^2 + c_1)I + \lambda_a T +
T^2 \right].
\label{eq:eiH_ml_fin_sum}
\end{equation}

The evolution operator for the neutrinos in the flavor basis is
thus given by
\begin{equation}
U_{f}(L) = e^{-i {\mathscr H}_{f} L} = U e^{-i {\mathscr H}_{m} L}
U^{-1} = \phi \sum_{a=1}^3 e^{-i L \lambda_a}
\frac{1}{3\lambda_a^2+c_1} \left[ (\lambda_a^2 + c_1)I + \lambda_a \tilde{T} +
\tilde{T}^2 \right],
\label{eq:evol}
\end{equation}
where $\tilde{T} \equiv U T U^{-1}$. Equation~(\ref{eq:evol}) is our
final expression for $U_{f}(L)$.

Let us pause for a moment to contemplate
Eqs.~(\ref{eq:eiH_ml_fin_sum}) and (\ref{eq:evol}). Since ${\mathscr
H}_f = U {\mathscr H}_{m} U^{-1}$, it is clear that $\tilde{T} = {\mathscr
H}_{f} - ({\rm tr \,} {\mathscr H}_{f}) I/3 = {\mathscr H}_{f} -
({\rm tr \,} {\mathscr H}_{m}) I/3$ due to the invariance of the
trace under transformation of $U$. In fact, the characteristic
equation is also invariant under $U$ and therefore so are the
coefficients $c_0$, $c_1$, $c_2$, and the eigenvalues
$\lambda_1$, $\lambda_2$, $\lambda_3$. However, the expression for
${\mathscr H}_{f}$ is much more complicated than that for ${\mathscr H}_{m}$,
which is the reason why we work with ${\mathscr H}_m$ instead of
${\mathscr H}_f$.

The formula~(\ref{eq:evol}) is our main result for the evolution
operator. It expresses the time (or $L$) evolution directly in terms
of the mass squared differences and the vacuum mixing angles without
introducing any auxiliary matter mixing angles.

\section{Probability amplitudes and transition probabilities}
\label{sec:prob}

The probability amplitude is defined as
\begin{equation}
A_{\alpha\beta} \equiv \langle \beta \vert U_f(L) \vert \alpha
\rangle, \quad \alpha,\beta=e,\mu,\tau.
\label{eq:ampl}
\end{equation}
Inserting Eq.~(\ref{eq:evol}) into Eq.~(\ref{eq:ampl}) gives
\begin{equation}
A_{\alpha\beta} = \phi \sum_{a=1}^3 e^{-i L \lambda_a}
\frac{(\lambda_a^2 + c_1)
\delta_{\alpha\beta} + \lambda_a \tilde{T}_{\alpha\beta} +
(\tilde{T}^2)_{\alpha\beta}}{3\lambda_a^2+c_1},
\label{eq:ampl_evol}
\end{equation}
where $\delta_{\alpha\beta}$ is Kronecker's delta and the entries
$\tilde{T}_{\alpha\beta}$ and
$(\tilde{T}^2)_{\alpha\beta}$, $\alpha,\beta=e,\mu,\tau$, are given in
Appendix~\ref{app:T_T2}. Here $\tilde{T}_{\alpha\beta} =
\tilde{T}_{\beta\alpha}$ and $(\tilde{T}^2)_{\alpha\beta} =
(\tilde{T}^2)_{\beta\alpha}$.

The probability of transition from a neutrino flavor
$\alpha$ to a neutrino flavor $\beta$ is thus defined by
the expression
\begin{equation}
P_{\alpha \beta} \equiv \left\vert A_{\alpha \beta} \right\vert^2 = A_{\alpha
\beta}^\ast A_{\alpha \beta}, \quad \alpha,\beta = e, \mu, \tau.
\end{equation}
Inserting the expression for the probability amplitude
into the definition of the transition probability, one finds
\begin{equation}
P_{\alpha\beta} = \sum_{a=1}^{3}\sum_{b=1}^{3} e^{-i L (\lambda_a -
\lambda_b)}
\frac{(\lambda_a^{2} + c_{1}) \delta_{\alpha\beta} + \lambda_a
\tilde{T}_{\alpha\beta} +
(\tilde{T}^2)_{\alpha\beta}}{3\lambda_{a}^{2}+c_{1}}
\frac{(\lambda_b^{2} + c_{1}) \delta_{\alpha\beta} + \lambda_b
\tilde{T}_{\alpha\beta} +
(\tilde{T}^2)_{\alpha\beta}}{3\lambda_{b}^{2}+c_{1}}.
\label{eq:prob}
\end{equation}
Since $\delta_{\alpha\beta}$, $\tilde{T}_{\alpha\beta}$, and
$(\tilde{T}^2)_{\alpha\beta}$ all are symmetric, it holds that
$P_{\alpha\beta} = P_{\beta\alpha}$.

Setting $A = 0$ gives
\begin{eqnarray}
&& \langle b \vert U_m(L) \vert a \rangle = \langle b \vert e^{-i
H_m L} \vert a \rangle = e^{-i E_a L} \delta_{ab}, \quad
a,b=1,2,3,\\
&& A_{\alpha\beta} = \langle \beta \vert U_f(L) \vert \alpha \rangle =
\sum_{a=1}^3 U_{\alpha a} U_{\beta a} e^{-i E_a L}, \quad
\alpha,\beta=e,\mu,\tau,
\end{eqnarray}
and
\begin{equation}
P_{\alpha\beta} = \delta_{\alpha\beta} - 4 \; \underset{a < b}{\sum_{a=1}^3
\sum_{b=1}^3} U_{\alpha a} U_{\beta a}
U_{\alpha b} U_{\beta b} \sin^2 x_{ab}, \quad
\alpha,\beta = e, \mu, \tau,
\label{eq:Pab_vac}
\end{equation}
where $x_{ab} \equiv E_{ab} L/2$, which are the well-known transition
probabilities for vacuum oscillations (see, e.g., Ref. \cite{ohls99}).

We can write the probabilities for oscillations in matter in a form
analogous to the ones for the vacuum probabilities, as
\begin{eqnarray}
P_{\alpha\beta} &=& \left(
\sum_{a=1}^{3}
\frac{(\lambda_a^2 + c_1) \delta_{\alpha\beta} + \lambda_a
\tilde{T}_{\alpha\beta} + (\tilde{T}^2)_{\alpha\beta}}{3 \lambda_a^2 +
c_1} \right)^2 \nonumber\\
&-& 4 \; \underset{a < b}{\sum_{a=1}^3
\sum_{b=1}^3} \frac{(\lambda_a^2 + c_1) \delta_{\alpha\beta} +
\lambda_a \tilde{T}_{\alpha\beta} +
(\tilde{T}^2)_{\alpha\beta}}{3 \lambda_a^2 + c_1}
\frac{(\lambda_b^2 + c_1) \delta_{\alpha\beta} + \lambda_b
\tilde{T}_{\alpha\beta} + (\tilde{T}^2)_{\alpha\beta}}{3 \lambda_b^2 + c_1}
\sin^{2}\tilde{x}_{ab}, \nonumber\\ \nonumber\\ && \alpha,\beta = e,\mu,\tau,
\label{eq:Pab_1}
\end{eqnarray}
where $\tilde{x}_{ab} \equiv (\lambda_a-\lambda_b) L/2$.
For $L=0$ Eq.~(\ref{eq:evol}) reads
\begin{equation}
I = \sum_{a=1}^3 \frac{1}{3 \lambda_a^2 + c_1} [ (\lambda_a^2 +
c_1) I + \lambda_a \tilde{T} + \tilde{T}^2 ].
\end{equation}
This means that the corresponding matrix element version of the above
formula is given by
\begin{equation}
\delta_{\alpha\beta} = \sum_{a=1}^3 \frac{(\lambda_a^2 +
c_1) \delta_{\alpha\beta} + \lambda_a \tilde{T}_{\alpha\beta} +
(\tilde{T}^2)_{\alpha\beta}}{3 \lambda_a^2 + c_1}, \quad \alpha,\beta =
e,\mu,\tau.
\end{equation}
Hence, the transition probabilities in matter are
\begin{eqnarray}
P_{\alpha\beta} &=& \delta_{\alpha\beta}
- 4 \; \underset{a < b}{\sum_{a=1}^3
\sum_{b=1}^3} \frac{(\lambda_a^2 + c_1) \delta_{\alpha\beta} +
\lambda_a \tilde{T}_{\alpha\beta} +
(\tilde{T}^2)_{\alpha\beta}}{3 \lambda_a^2 + c_1}
\frac{(\lambda_b^2 + c_1) \delta_{\alpha\beta} + \lambda_b
\tilde{T}_{\alpha\beta} + (\tilde{T}^2)_{\alpha\beta}}{3 \lambda_b^2 + c_1}
\sin^{2}\tilde{x}_{ab}, \nonumber\\ \nonumber\\
&& \alpha,\beta = e,\mu,\tau.
\label{eq:Pab_2}
\end{eqnarray}

From unitarity, there are only three independent transition
probabilities, since the other three can be obtained from them,
i.e., from the equations
\begin{eqnarray}
&& P_{ee} + P_{e\mu} + P_{e\tau} = 1, \\
&& P_{\mu e} + P_{\mu\mu} + P_{\mu\tau} = 1, \\
&& P_{\tau e} + P_{\tau\mu} + P_{\tau\tau} = 1,
\end{eqnarray}
where $P_{e\mu} = P_{\mu e}$, $P_{e\tau} = P_{\tau e}$, and
$P_{\mu\tau} = P_{\tau\mu}$. We will choose $P_{e\mu}$, $P_{e\tau}$,
and $P_{\mu\tau}$ as the three independent ones. The expression for
the transition probabilities in Eq.~(\ref{eq:Pab_2}) thus simplifies to
\begin{equation}
P_{\alpha\beta}=
- 4 \; \underset{a < b}{\sum_{a=1}^3
\sum_{b=1}^3} \frac{\lambda_a \tilde{T}_{\alpha\beta} +
(\tilde{T}^2)_{\alpha\beta}}{3 \lambda_a^2 + c_1}
\frac{\lambda_b \tilde{T}_{\alpha\beta} +
(\tilde{T}^2)_{\alpha\beta}}{3 \lambda_b^2 + c_1}
\sin^{2}\tilde{x}_{ab},
\label{eq:Pab_n}
\end{equation}
which is valid only for $\alpha \neq \beta$.

\section{Determining the auxiliary matter mixing angles}
\label{sec:mma}

The flavor states can be expressed as linear combinations of either the
vacuum mass eigenstates ($A = 0$) in the basis ${\cal H}_m$ or the
matter mass eigenstates ($A \neq 0$) in the basis ${\cal H}_M$. The
corresponding components are related to each other as follows:
\begin{eqnarray}
\psi_f &=& U \psi_m,\\
\psi_f &=& U^M \psi_M,
\end{eqnarray}
where $U = U(\theta_1,\theta_2,\theta_3)$ and $U^M =
U^M(\theta_1^M, \theta_2^M, \theta_3^M)$ are the unitary mixing
matrices for vacuum and matter, respectively. Here $\theta^M_i$,
$i = 1,2,3$, are the auxiliary (effective) matter mixing angles.

Combining these expressions for the flavor components, one obtains the
following relation between the two different sets of mass eigenstate
components:
\begin{equation}
\psi_M = R \psi_m,
\end{equation}
where
$$
R \equiv (U^M)^{-1} U.
$$
The matrix $R$ is, of course, a unitary matrix (even orthogonal,
since $U$ and $U^M$ are real). This means that the matter mixing
matrix can be expressed in the vacuum mixing matrix as
\begin{equation}
U^M(\theta_1^M, \theta_2^M, \theta_3^M) =
U(\theta_1,\theta_2,\theta_3) R^{-1}.
\label{eq:UUR}
\end{equation}
The relations between the different bases can be depicted as in the
following commutative diagram:

\begin{picture}(12,6)
\put(4,4){$\psi_m \in {\cal H}_m$}
\put(6,4.1){\vector(1,0){1.75}}
\put(8.25,4){${\cal H}_f \ni \psi_f$}
\put(5.25,3.5){\vector(0,-1){1.5}}
\put(4,1.25){$\psi_{M} \in {\cal H}_M$}
\put(6,2){\vector(4,3){1.9}}
\put(6.75,4.5){$U$}
\put(4.6,2.65){$R$}
\put(7.15,2.15){$U^M$}
\end{picture}

From this diagram one readily obtains the Hamiltonian in presence of
matter ${\mathscr H}_M$ as
$$
{\mathscr H}_M \equiv R {\mathscr H}_m R^{-1} \quad \mbox{or}
\quad {\mathscr H}_M \equiv R T R^{-1} + \tfrac{1}{3} ({\rm tr \,}
{\mathscr H}_m) I,
$$
which is diagonal in the mass eigenstate basis in matter, ${\cal H}_{M}$.

Due to the invariance of the trace, we have
\begin{equation}
T_M \equiv {\mathscr H}_M - \tfrac{1}{3} ({\rm tr \,} {\mathscr H}_M)
I = R T R^{-1},
\label{eq:tm}
\end{equation}
where $T_M$ is a traceless diagonal matrix. This implies that
\begin{equation}
e^{-i L T_M} = R e^{-i L T} R^{-1} = \sum_{a=1}^{3} e^{-iL\lambda_{a}}
\frac{1}{3\lambda_{a}^{2}+c_{1}} \left[ (\lambda_{a}^{2}+c_{1})I
+\lambda_{a}T_{M}+T_{M}^{2} \right],
\label{eq:exptm}
\end{equation}
which is actually equivalent to the system of
equations~(\ref{eq:eqsys}). 

Inverting the first relation above, we obtain
\begin{equation}
U_f(L) = \phi U e^{-i L T} U^{-1} = \phi U R^{-1} e^{-i L T_M} R U^{-1}
= \phi U^M e^{-i L T_M} ({U^M})^{-1}.
\label{eq:exptm2}
\end{equation}
From this one immediately obtains the transition amplitude
\begin{equation}
A_{\alpha\beta} = \langle \beta|U_{f}(L)|\alpha \rangle =
\phi \sum_{a=1}^{3} U^M_{\alpha a} U^M_{\beta a} e^{-iL\lambda_{a}},
\quad \alpha,\beta=e,\mu,\tau,
\label{eq:ampl2}
\end{equation}
where the $U^M_{\alpha a}$'s are the entries in the matrix $U^M$.
This allows us to identify the $U^M$ matrix elements with those in
Eq.~(\ref{eq:ampl_evol}) as
\begin{equation}
U^M_{\alpha a} U^M_{\beta a} = \frac{(\lambda_a^2 + c_1) \delta_{\alpha\beta} +
\lambda_a \tilde{T}_{\alpha\beta} + (\tilde{T}^2)_{\alpha\beta}}{3
\lambda_a^2 + c_1}, \quad \alpha,\beta = e,\mu,\tau, \quad a=1,2,3.
\label{eq:uum1}
\end{equation}
The matter mixing angles can therefore be expressed as follows
\begin{equation}
\theta^M_1 \equiv \arctan \frac{U^M_{\mu
3}}{U^M_{\tau 3}} = \arctan \frac{\lambda_3 \tilde{T}_{e\mu} +
(\tilde{T}^2)_{e\mu}}{\lambda_3 \tilde{T}_{e\tau} +
(\tilde{T}^2)_{e\tau}}, \label{mm1}
\end{equation}
\begin{equation}
\theta^M_2 \equiv \arcsin U^M_{e 3} = \arcsin \sqrt{\frac{\lambda_3^2 +
c_1 + \lambda_3 \tilde{T}_{ee} + (\tilde{T}^2)_{ee}}{3 \lambda_3^2 +
c_1}}, \label{mm2}
\end{equation}
\begin{equation}
\theta^M_3 \equiv \arctan
\frac{U^M_{e 2}}{U^M_{e 1}} = \arctan \frac{\lambda_2 \tilde{T}_{e\alpha} +
(\tilde{T}^2)_{e\alpha}}{\lambda_1 \tilde{T}_{e\alpha} +
(\tilde{T}^2)_{e\alpha}} \frac{3 \lambda_1^2 + c_1}{3 \lambda_2^2 +
c_1}, \quad \alpha = \mu,\tau. \label{mm3}
\end{equation}
The transition probabilities in matter can thus be written as
\begin{equation}
P_{\alpha\beta} = \delta_{\alpha\beta} - 4 \; \underset{a < b}{\sum_{a=1}^3
\sum_{b=1}^3} U^M_{\alpha a} U^M_{\beta a}
U^M_{\alpha b} U^M_{\beta b} \sin^2 \tilde{x}_{ab}, \quad
\alpha,\beta = e, \mu, \tau.
\label{eq:Pab_3}
\end{equation}
The above formulas for the transition probabilities are thus
in effect exactly the same as the formulas in Eq.~(\ref{eq:Pab_2}).

When the matter density goes to zero, i.e., $A \to 0$, then
$U^M_{\alpha a} \to U_{\alpha a}$ ($\theta^M_i \to \theta_i$) and
$\tilde{x}_{ab} \to x_{ab}$. Also Eqs.~(\ref{eq:Pab_2}) and
(\ref{eq:Pab_3}) will become identical to Eq.~(\ref{eq:Pab_vac}) when $A =
0$.

To connect this treatment to the usual treatment in terms of the
auxiliary matter mixing angles, we have to find the matrix $R$.
Using Eq.~(\ref{eq:tm}), the matrix $R$, which diagonalizes the matrix
$T$, can be constructed from the eigenvectors of the matrix $T$ as
\begin{equation}
R = \left( \begin{array}{c} {\bf v}_1^T \\ {\bf v}_2^T \\ {\bf v}_3^T
\end{array} \right),
\end{equation}
i.e.,
\begin{equation}
R^{-1} = R^T = \left( \begin{array}{ccc} {\bf v}_1 & {\bf v}_2 &
{\bf v}_3 \end{array} \right),
\end{equation}
where
$$
{\bf v}_a = \left( \begin{array}{c} \alpha_a \\ \beta_a \\ 1
\end{array} \right) z_a, \quad a = 1,2,3.
$$
Here
$$
\alpha_a = \frac{T_{12} \left( T_{33} -
\lambda_a \right) - T_{13} T_{23}}{T_{23} \left( T_{11} - \lambda_a
\right) - T_{12} T_{13}}, \quad \beta_a = \frac{T_{12} \left( T_{33} -
\lambda_a \right) - T_{13} T_{23}}{T_{13} \left( T_{22} - \lambda_a
\right) - T_{12} T_{23}}, \quad z_a = \frac{1}{\sqrt{\alpha_a^2 + \beta_a^2
+ 1}}, \quad a = 1,2,3.
$$
The orthogonal matrix $R$ is of course diagonalizing both $T$ and
${\mathscr H}_m$.

Equation~(\ref{eq:UUR}) can now be written in matrix form as
\begin{eqnarray}
\left( \begin{array}{ccc} U^M_{e 1} & U^M_{e 2} &
U^M_{e 3} \\ U^M_{\mu 1} & U^M_{\mu 2} & U^M_{\mu 3} \\ U^M_{\tau 1} &
U^M_{\tau 2} & U^M_{\tau 3} \end{array} \right) &\equiv&
\left( \begin{array}{ccc} C^M_2 C^M_3 & S^M_3
C^M_2 & S^M_2 \\ - S^M_3 C^M_1 -
S^M_1 S^M_2 C^M_3 & C^M_1 C^M_3 -
S^M_1 S^M_2 S^M_3 & S^M_1 C^M_2 \\
S^M_1 S^M_3 - S^M_2
C^M_1 C^M_3 & - S^M_1 C^M_3 - S^M_2
S^M_3 C^M_1 & C^M_1 C^M_2 \end{array} \right) \nonumber\\
&=&
\left( \begin{array}{ccc} C_2 C_3 & S_3 C_2 & S_2 \\ - S_3 C_1 -
S_1 S_2 C_3 & C_1 C_3 - S_1 S_2 S_3 & S_1 C_2 \\ S_1 S_3 - S_2
C_1 C_3 & - S_1 C_3 - S_2 S_3 C_1 & C_1 C_2 \end{array} \right)
\nonumber\\
&\times& \left( \begin{array}{ccc} \alpha_1 z_1 & \alpha_2 z_2 &
\alpha_3 z_3 \\ \beta_1 z_1 & \beta_2 z_2 & \beta_3 z_3 \\ z_1 & z_2 & z_3
\end{array} \right) \nonumber\\
&\equiv& \left( \begin{array}{ccc} U_{e 1} & U_{e 2} &
U_{e 3} \\ U_{\mu 1} & U_{\mu 2} & U_{\mu 3} \\ U_{\tau 1} & U_{\tau
2} & U_{\tau 3} \end{array} \right) \left( \begin{array}{ccc} ({\bf
v}_1)_1 & ({\bf v}_2)_1 & ({\bf v}_3)_1 \\ ({\bf
v}_1)_2 & ({\bf v}_2)_2 & ({\bf v}_3)_2 \\ ({\bf
v}_1)_3 & ({\bf v}_2)_3 & ({\bf v}_3)_3 \end{array} \right),
\end{eqnarray}
where $S^M_i \equiv \sin \theta^M_i$ and $C^M_i \equiv \cos
\theta^M_i$ for $i = 1,2,3$.
We can then obtain the matter mixing angles from the
above matrix relation as
\begin{equation}
\theta_1^M = \arctan \frac{S^M_1}{C^M_1} = \arctan \frac{(-S_3 C_1
- S_1 S_2 C_3) \alpha_3 z_3 + (C_1 C_3 - S_1 S_2 S_3) \beta_3 z_3 +
S_1 C_2 z_3}{(S_1 S_3 - S_2 C_1 C_3) \alpha_3 z_3 + (-S_1 C_3 - S_2
S_3 C_1) \beta_3 z_3 + C_1 C_2 z_3},
\label{eq:tM1}
\end{equation}
\begin{equation}
\theta_2^M = \arcsin S^M_2 = \arcsin (C_2
C_3 \alpha_3 z_3 + S_3 C_2 \beta_3 z_3 + S_2 z_3),
\end{equation}
\begin{equation}
\theta_3^M = \arctan \frac{S^M_3}{C^M_3} = \arctan \frac{C_2 C_3
\alpha_2 z_2 + S_3 C_2 \beta_2 z_2 + S_2 z_2}{C_2 C_3 \alpha_1
z_1 + S_3 C_2 \beta_1 z_1 + S_2 z_1}. \label{eq:tM3}
\end{equation}

The conditions for resonance are given by the extremal points
of the probability amplitudes. These will coincide with extremal points for the
modulus of the amplitudes, thus for the products $U^M_{\alpha
a} U^M_{\beta a}$ or for the expressions in the right-hand side of
Eq.~(\ref{eq:uum1}) as functions of $A$.

\section{Discussion}
\label{sec:disc}

The main result of our analysis is given by the time evolution
operator for the neutrinos when passing through matter with constant matter
density in Eq.~(\ref{eq:evol}) expressed as a finite sum of elementary
functions in the matrix elements of the Hamiltonian.
Also the transition probabilities in Eq.~(\ref{eq:Pab_n})
and the neutrino energies in presence of matter $\tilde{E}_a = \lambda_a$,
where $a=1,2,3$, given by Eqs.~(\ref{eq:l1})--(\ref{eq:l3}) belong to our main
results as do the expressions for the matter mixing
angles in Eqs.~(\ref{mm1})--(\ref{mm3}). In our treatment these
auxiliary matter mixing angles play no independent role and are not
really needed. It is convenient, though, to express the matter mixing
angles in the matrix elements of the Hamiltonian to look for the
resonance conditions.

\begin{figure}
\begin{center}
\epsfig{figure=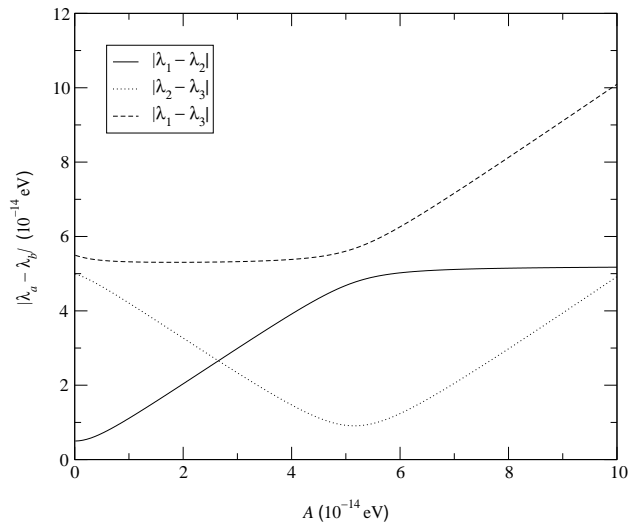,height=12cm,angle=-90,clip=on}
\caption{The differences $\vert \lambda_a - \lambda_b \vert$,
$a,b=1,2,3$, $a \neq b$, as a function of the matter density $A$ for
$\theta_1 = \theta_3 = 45^\circ$ (bimaximal mixing), $\theta_2 = 5^\circ$,
$\Delta m^2 = 10^{-4} \; {\rm eV}^2$, $\Delta M^2 =
10^{-3} \; {\rm eV}^2$, and $E = 10 \; {\rm GeV}$.}
\label{fig:1}
\end{center}
\end{figure}

\begin{figure}
\begin{center}
\epsfig{figure=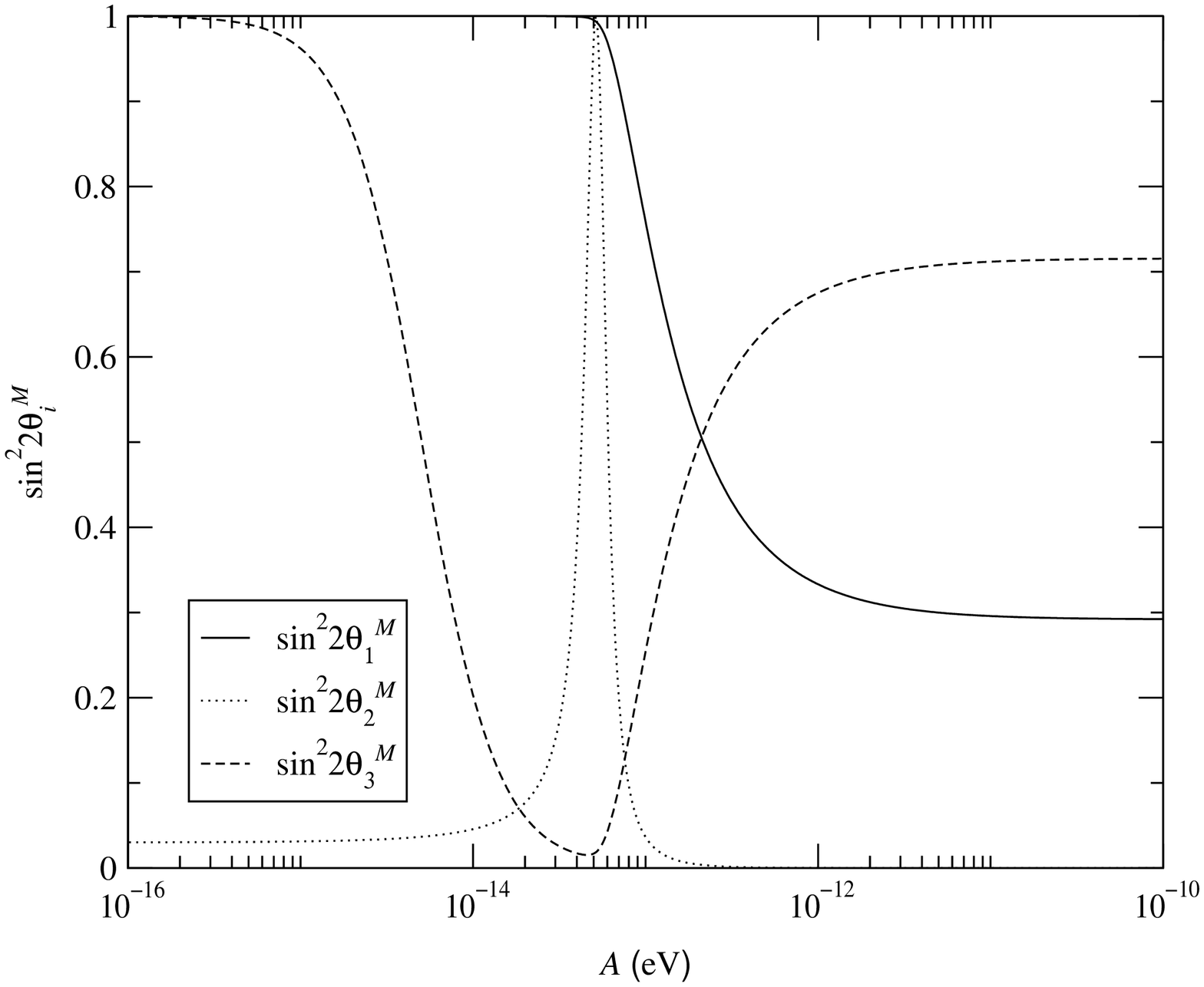,height=12cm,angle=-90,clip=on}
\caption{The quantities $\sin^2 2\theta^M_i$, $i=1,2,3$, as a function
of the matter density $A$ for
$\theta_1 = \theta_3 = 45^\circ$ (bimaximal mixing), $\theta_2 = 5^\circ$,
$\Delta m^2 = 10^{-4} \; {\rm eV}^2$, $\Delta M^2 =
10^{-3} \; {\rm eV}^2$, and $E = 10 \; {\rm
GeV}$. Note that $\sin^2 2\theta^M_i = 1$ corresponds to maximal mixing.}
\label{fig:2}
\end{center}
\end{figure}

\begin{figure}
\begin{center}
\epsfig{figure=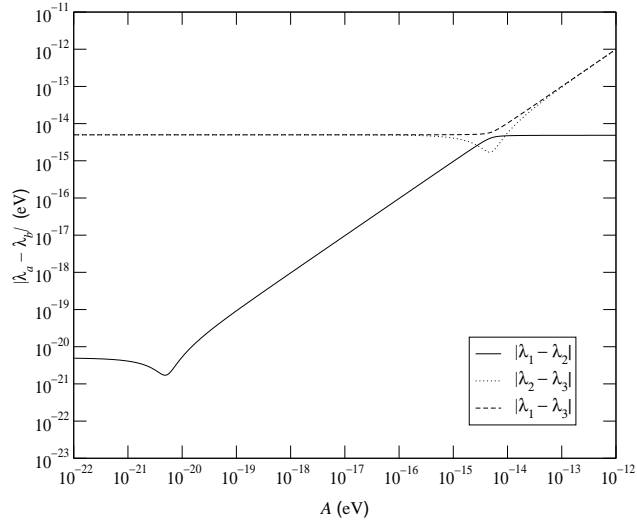,height=12cm,angle=-90,clip=on}
\caption{The differences $\vert \lambda_a - \lambda_b \vert$,
$a,b=1,2,3$, $a \neq b$, as a function of the matter density $A$ for
$\theta_1 = \theta_2 = \theta_3 = 10^\circ$,
$\Delta m^2 = 10^{-10} \; {\rm eV}^2$, $\Delta M^2 =
10^{-4} \; {\rm eV}^2$, and $E = 10 \; {\rm GeV}$.}
\label{fig:3}
\end{center}
\end{figure}

\begin{figure}
\begin{center}
\epsfig{figure=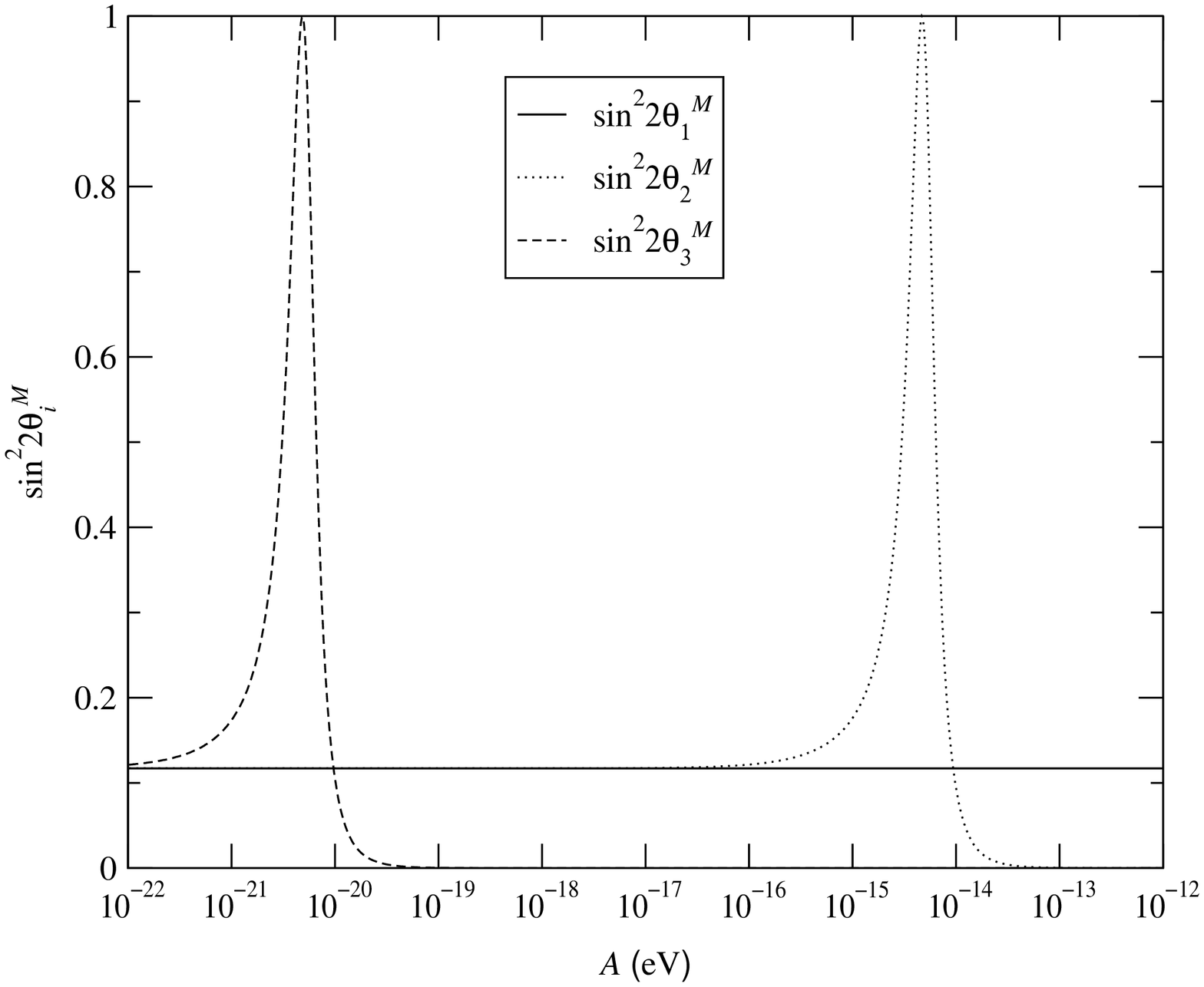,height=12cm,angle=-90,clip=on}
\caption{The quantities $\sin^2 2\theta^M_i$, $i=1,2,3$, as a function
of the matter density $A$ for $\theta_1 = \theta_2 = \theta_3 = 10^\circ$,
$\Delta m^2 = 10^{-10} \; {\rm eV}^2$, $\Delta M^2 =
10^{-4} \; {\rm eV}^2$, and $E = 10 \; {\rm
GeV}$. Note that $\sin^2 2\theta^M_i = 1$ corresponds to maximal mixing.}
\label{fig:4}
\end{center}
\end{figure}

As an illustration of the resonance phenomena, we have plotted the
energy differences $\vert \lambda_{a} - \lambda_{b} \vert$, where $a,b
= 1,2,3$, $a \neq b$, and the behavior of the matter mixing angles
$\theta_{1}^{M}$, $\theta_{2}^{M}$, $\theta_{3}^{M}$ as a function of
the matter density $A$ for two different cases. First for two mass squared
differences that are close and bimaximal mixing \cite{barg98},
i.e., two large mixing angles and one small. The results are
presented in Figs.~\ref{fig:1} and \ref{fig:2}. Then for two widely
separated mass squared differences and all the mixing angles
small. These results are presented in Figs.~\ref{fig:3} and
\ref{fig:4}. The energy differences $\vert \lambda_a - \lambda_b
\vert$ are related to the effective matter mass squared differences
$\Delta \tilde{m}_{ab}^2$ by the following relations
$$
\vert \Delta \tilde{m}_{ab}^2 \vert = 2E \vert \lambda_a -
\lambda_b \vert, \qquad a,b=1,2,3, \quad a \neq b.
$$

As can be seen from Figs.~\ref{fig:1}--\ref{fig:4}, the resonances occur
(i.e., the mixing is maximal) when the energy levels in presence
of matter approach each other as function of the matter density $A$.
As the mass squared differences approach each other in magnitude, the
resonances move closer together and the mixing angle $\theta^{M}_{1}$, which
looks quite flat (constant) at large separations of the mass squared
differences, will have a modest increase above the location of the
(higher) resonance of the mixing angle $\theta^{M}_{2}$. The smaller
the vacuum mixing angles are, the shaper are the resonance peaks. For
bimaximal mixing the resonance occurs only for the small mixing angle, but
even for matter densities above the resonance region the values of the
auxiliary mixing angles are changed appreciably from their vacuum
values, as is shown in Fig.~\ref{fig:2}.

Our results for the auxiliary matter mixing angles are in agreement with
the previous approximative and numerical calculations in
Refs. \cite{kim87,zagl88,josh88,tosh87,aqui95}.

\begin{figure}
\begin{center}
\epsfig{figure=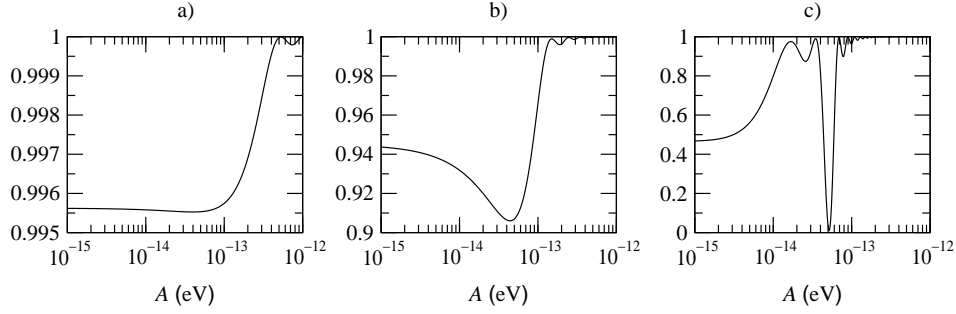,width=5cm,angle=-90,clip=on}
\caption{The transition probability $P_{ee}$ as a function of the
matter density $A$ for $\theta_1 = \theta_3 = 45^\circ$ (bimaximal
mixing), $\theta_2 = 5^\circ$, $\Delta m^2 = 10^{-4} \; {\rm eV}^2$,
$\Delta M^2 = 10^{-3} \; {\rm eV}^2$, and $L/E = \eta
(2R_\oplus / 10 \; {\rm GeV}) \simeq 3.23 \times 10^4 \; {\rm
eV}^{-2} \; \eta$, where $R_\oplus \simeq 6378 \; {\rm km} \simeq 1.62
\times 10^{14} \; {\rm eV^{-1}}$ is the
(equatorial) radius of the Earth. (a) $\eta = 1/5$, (b) $\eta = 1$, and
(c) $\eta = 5$.}
\label{fig:5}
\end{center}
\end{figure}

\begin{figure}
\begin{center}
\epsfig{figure=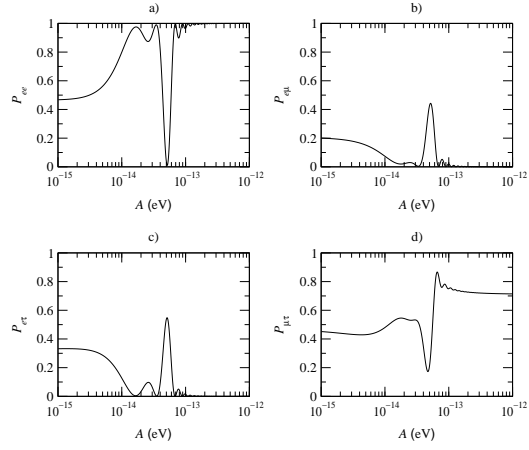,height=9cm,angle=-90,clip=on}
\caption{Transition probabilities as functions of the matter density
$A$, using the same parameter values as
for Fig.~\ref{fig:5}(c). (a) $P_{ee}$, (b) $P_{e\mu}$, (c) $P_{e\tau}$,
and (d) $P_{\mu\tau}$.}
\label{fig:6}
\end{center}
\end{figure}

We have also calculated the transition probabilities $P_{\alpha\beta}$ for
neutrino oscillations in matter as functions of the matter density $A$. The
results are illustrated in Figs.~\ref{fig:5} and \ref{fig:6}. These two
figures correspond to the only physically measurable quantities. Our
illustration is chosen to correspond to values of the mass squared
differences that are close to those obtained from analyses of
atmospheric neutrino data. In Fig.~\ref{fig:5}(c), we can see a sharp
drop in the probability $P_{ee}$ at a matter density of about $5 \cdot
10^{-14} \; {\rm eV}$ (same parameter values as in Figs.~\ref{fig:1}
and \ref{fig:2}) for a sensitivity of $L/E \approx 1.6 \cdot
10^{5} \; {\rm eV}^{-2}$ as the neutrinos pass through the Earth. This
corresponds to an enhancement of $P_{e\mu}$ and $P_{e\tau}$ at the
same value of $A$, whereas the probability $P_{\mu\tau}$ wiggles in
the same region. Due to unitarity only three of these probabilities
are of course independent.

The Cayley--Hamilton formalism could in principle be used to derive
neutrino oscillation formulas in matter for arbitrary numbers of neutrino
flavors. However, the calculations for four or even more flavors will be
much more tedious than in the case of three neutrino flavors. For four
or more flavors, the unitary mixing matrices have more complicated structures
than in the three flavor case. Another complication, for more than
four flavors, is that one will not, in general, be able to find analytical
solutions to the characteristic equation, since only characteristic
equations up to degree four are analytically solvable.

In Appendix~\ref{app:twonu}, we have given, as a comparison and
reference, the derivation of the neutrino oscillation formulas in
matter for two neutrino flavors also using the Cayley--Hamilton
formalism. We observe that the derivation for two flavors is much
easier to make than that for three flavors. The evolution
operator $U_m(L)$ can of course be written in a much more compact form for two
flavors than for three; compare Eqs.~(\ref{eq:eiH_ml_fin_sum}) and
(\ref{eq:evol_two}).

\acknowledgments

We would like to thank G{\"o}ran Lindblad for useful discussions.
This work was supported by the Swedish Natural Science Research
Council (NFR), Contract No. F-AA/FU03281-312. Support for this work
was also provided by the Engineer Ernst Johnson Foundation (T.O.).

\appendix

\section{Introducing the relativistic limit}
\label{app:rel}

We will here introduce the relativistic limit by reparametrizing the
expressions for the energy differences. Any two of $E_{21}$, $E_{32}$,
and $E_{13}$
can be chosen as independent variables. Thus, we use $E_{ab} =
\Delta m_{ab}^{2}/(E_{a}+E_{b})$ for $a,b=1,2,3$ and $a \neq b$, where
$\Delta m_{ab}^2 \equiv m_{a}^{2}-m_{b}^{2}$. Without lack of
generality, we will assume that there is a mass ordering with $m_{1}
< m_{2} < m_{3}$. Apart from the masses, we will take the momentum as
a common independent variable and assume that the momentum is the same
for all components of the mixed state. However, it is convenient to
use instead the energy as an independent quantity. This choice can
again be made in several ways. For definiteness, we will here choose as
physical variables $E \equiv (E_{1} + E_{2})/2$,
$\Delta m^{2} \equiv \Delta m_{21}^{2}$, and $\Delta M^{2} \equiv
\Delta m_{32}^{2}$. This gives
\begin{equation}
E_{21} = \frac{\Delta m^{2}}{2E},\label{eq:dm2}
\end{equation}
\begin{equation}
E_{32} = - \left(E+\frac{\Delta m^{2}}{2E}\right) +
\sqrt{\left(E+\frac{\Delta m^{2}}{2E}\right)^{2} + \Delta M^{2}}.
\end{equation}

Since in the applications $\Delta M/E \ll 1$ ($\lesssim 10^{-6}$), we
find the excellent approximation
\begin{equation}
E_{32} \simeq \frac{\Delta M^{2}}{2E}.
\label{eq:Dm2}
\end{equation}
Equations~(\ref{eq:dm2}) and (\ref{eq:Dm2}) lead to the following
expressions for the coefficients $c_{2}$, $c_{1}$, and $c_{0}$:
\begin{equation}
c_2 = 0,
\end{equation}
\begin{eqnarray}
c_1 &\simeq& - \frac{1}{3} A^2 + \frac{A}{6E} \left[ U_{e1}^2 \left(
\Delta M^{2} + 2 \Delta m^{2} \right) + U_{e2}^2 \left( \Delta M^{2} -
\Delta m^{2} \right) - U_{e3}^2 \left( 2 \Delta M^{2} + \Delta m^{2} \right)
\right] \nonumber\\
&-& \frac{1}{12 E^{2}} \left(\Delta M^{4} + \Delta m^{4} +
\Delta M^{2} \Delta m^{2} \right),
\end{eqnarray}
\begin{eqnarray}
c_0 &\simeq& - \frac{2}{27} A^3 + \frac{A^2}{18E} \left[ U_{e1}^2
\left( \Delta M^{2} + 2 \Delta m^{2} \right) + U_{e2}^2 \left( \Delta
M^{2} - \Delta m^{2} \right) - U_{e3}^2 \left( 2 \Delta M^{2} + \Delta
m^{2} \right) \right] \nonumber\\
&+& \frac{A}{36 E^{2}} \big[ U_{e1}^2 \left(2 \Delta M^{2} + \Delta
m^{2} \right) \left( \Delta M^{2} - \Delta m^{2} \right) + U_{e2}^2
\left( 2 \Delta M^{2} + \Delta m^{2} \right) \left( \Delta M^{2} + 2
\Delta m^{2} \right) \nonumber\\
&-& U_{e3}^2 \left( \Delta M^{2} + 2 \Delta m^{2}
\right) \left( \Delta M^{2} - \Delta m^{2} \right) -
\left( \Delta M^{4} + \Delta m^{4} + \Delta M^{2} \Delta m^{2}
\right) \big] \nonumber\\
&-& \frac{1}{216 E^{3}} \left( 2 \Delta M^{2} + \Delta m^{2} \right)
\left( \Delta M^{2} + 2 \Delta m^{2} \right) \left( \Delta M^{2} -
\Delta m^{2} \right).
\end{eqnarray}

\section{The entries of the matrices $\tilde{T}$ and $\tilde{T}^2$}
\label{app:T_T2}

The entries of the $3 \times 3$ real symmetric matrix $T^2$ are
\begin{equation}
(T^2)_{11} = \tfrac{1}{3} \left[ A^2 \left( U_{e1}^2 + \tfrac{1}{3}
\right) + 2A
\left( U_{e1}^2 - \tfrac{1}{3} \right) \left( E_{12} + E_{13} \right)
+ \tfrac{1}{3} \left(E_{12} + E_{13}\right)^2 \right],
\end{equation}
\begin{equation}
(T^2)_{22} = \tfrac{1}{3} \left[ A^2 \left( U_{e2}^2 + \tfrac{1}{3}
\right) + 2A
\left( U_{e2}^2 - \tfrac{1}{3} \right) \left( E_{21} + E_{23} \right)
+ \tfrac{1}{3} \left(E_{21} + E_{23}\right)^2 \right],
\end{equation}
\begin{equation}
(T^2)_{33} = \tfrac{1}{3} \left[ A^2 \left( U_{e3}^2 + \tfrac{1}{3}
\right) + 2A
\left( U_{e3}^2 - \tfrac{1}{3} \right) \left( E_{31} + E_{32} \right)
+ \tfrac{1}{3} \left(E_{31} + E_{32}\right)^2 \right],
\end{equation}
\begin{equation}
(T^2)_{12} = (T^2)_{21} = \tfrac{1}{3} U_{e1} U_{e2} A \left( A +
E_{13} + E_{23} \right),
\end{equation}
\begin{equation}
(T^2)_{13} = (T^2)_{31} = \tfrac{1}{3} U_{e1} U_{e3} A \left( A +
E_{12} + E_{32} \right),
\end{equation}
\begin{equation}
(T^2)_{23} = (T^2)_{32} = \tfrac{1}{3} U_{e2} U_{e3} A \left( A +
E_{21} + E_{31} \right).
\end{equation}
For the matrices $\tilde{T}$ and $\tilde{T}^2$ we have the following
expressions for the entries:
\begin{equation}
\tilde{T}_{\alpha \beta} = \sum_{a=1}^3 \sum_{b=1}^3 U_{\alpha a}
U_{\beta b} T_{ab}, \quad \alpha,\beta = e,\mu,\tau,
\label{eq:t}
\end{equation}
and
\begin{equation}
(\tilde{T}^{2})_{\alpha\beta} = \sum_{a=1}^3 \sum_{b=1}^3 U_{\alpha a}
U_{\beta b} (T^2)_{ab}, \quad \alpha,\beta = e,\mu,\tau.
\label{eq:t2}
\end{equation}
For $\alpha \neq \beta$ Eq.~(\ref{eq:t}) simplifies to
\begin{equation}
\tilde{T}_{\alpha \beta} = \sum_{a=1}^3 U_{\alpha a} U_{\beta a} E_a.
\label{eq:t1}
\end{equation}

\section{The two flavor neutrino case}
\label{app:twonu}

Considering only a two flavor neutrino oscillation model instead of a
three neutrino, the free Hamiltonian in the mass eigenstate basis and
the potential matrix are
\begin{equation}
H_m = \left( \begin{array}{cc} E_1 & 0 \\ 0 & E_2 \end{array} \right)
\quad \mbox{and} \quad V_f = \left( \begin{array}{cc} A & 0 \\ 0 & 0
\end{array} \right).
\end{equation}
In the mass eigenstate basis, the potential matrix is given by
\begin{equation}
V_m = U^{-1} V_f U,
\end{equation}
where
$$
U = \left( \begin{array}{cc} U_{e1} & U_{e2} \\ U_{\mu 1} & U_{\mu 2}
\end{array} \right) = \left( \begin{array}{cc} \cos \theta & \sin
\theta \\ - \sin \theta & \cos \theta \end{array} \right)
$$
is the usual $2 \times 2$ orthogonal matrix, which describes neutrino
mixing with two neutrino flavors. Thus, we have
\begin{equation}
V_m = A \left( \begin{array}{cc} U_{e1}^2 & U_{e1} U_{e2} \\ U_{e1}
U_{e2} & U_{e2}^2 \end{array} \right)
\end{equation}
and
\begin{equation}
{\mathscr H}_m = H_m + V_m = \left( \begin{array}{cc} E_1 + A U_{e1}^2
& A U_{e1} U_{e2} \\ A U_{e1} U_{e2} & E_2 + A U_{e2}^2 \end{array}
\right).
\end{equation}
Again, we have to find the evolution operator $U_m(L) = e^{-i {\mathscr H}_m
L}$. Similar but simpler formulas than in the three flavor neutrino
case now give
\begin{equation}
U_m(L) = e^{-i {\mathscr H}_m L} = \phi e^{-i L T} = \phi \sum_{a=1}^2
e^{-i L \lambda_a} \frac{1}{2 \lambda_a} \left( \lambda_a I + T
\right) = \phi \left( \cos (L
\tilde{\lambda})  - i \frac{1}{\tilde{\lambda}} \sin (L
\tilde{\lambda}) T \right),
\label{eq:evol_two}
\end{equation}
where $\phi \equiv e^{-i L {\rm tr\,} {\mathscr H}_m/2}$, $T \equiv
{\mathscr H}_m - ({\rm tr \,}{\mathscr H}_m) I/2$, $\lambda_1 =
- \tilde{\lambda}$, $\lambda_2 = \tilde{\lambda}$, and
$\tilde{\lambda} = \sqrt{-\det T}$.
Here
$$
\tilde{\lambda} \equiv \sqrt{\frac{1}{4} A^2 - \frac{x_{21}}{L} \left(
A \cos 2\theta - \frac{x_{21}}{L} \right)}, \quad x_{21} \equiv \frac{E_{21}
L}{2} = \frac{\Delta m^2 L}{4E}, \quad \Delta m^2 \equiv m_2^2 - m_1^2.
$$
The quantity $\tilde{\lambda}$ comes from the solutions, $\lambda_1$
and $\lambda_2$, to the characteristic equation $\lambda^2 + c_1
\lambda + c_0 = 0$, where $c_1 = - {\rm tr \,}T = 0$ and $c_0 = \det T
= - A^2/4 + A U_{e1}^2 E_{21}/2 + A U_{e2}^2
E_{12}/2 + E_{12} E_{21}/4 = - A^2/4 +
x_{21} \left( A \cos 2\theta - x_{21}/L \right)/L$.

The evolution operator $U_m(L)$ can now be written in matrix form as
\begin{equation}
U_m(L) = \phi \left( \begin{array}{cc} \alpha & \beta \\ \beta &
\alpha^\ast \end{array} \right),
\end{equation}
where $\alpha = \cos (L \tilde{\lambda}) - i \left( A \cos 2\theta - 2
x_{21}/L \right) \sin (L \tilde{\lambda})/(2 \tilde{\lambda})$ and
$\beta = - i A \sin 2\theta \sin (L \tilde{\lambda})/(2 \tilde{\lambda})$.
Similarly, the evolution operator $U_f(L)$ is
\begin{eqnarray}
U_f(L) &=& U U_m(L) U^{-1} \nonumber\\
&=& \phi \left( \begin{array}{cc} \cos \theta &
\sin \theta \\ - \sin \theta & \cos \theta \end{array} \right) \left(
\begin{array}{cc} \alpha & \beta \\ \beta & \alpha^\ast \end{array} \right)
\left( \begin{array}{cc} \cos \theta & - \sin \theta \\ \sin \theta &
\cos \theta \end{array} \right) \nonumber \\
&=& \phi \left( \begin{array}{cc} \alpha
\cos^2 \theta + \alpha^\ast \sin^2 \theta + \beta \sin 2\theta &
\frac{\alpha^\ast - \alpha}{2} \sin 2\theta + \beta \cos 2\theta \\
\frac{\alpha^\ast - \alpha}{2} \sin 2\theta + \beta \cos 2\theta &
\alpha \sin^2 \theta + \alpha^\ast \cos^2 \theta - \beta \sin 2\theta
\end{array} \right).
\end{eqnarray}
The $e\mu$-element of the $U_f(L)$ matrix is
\begin{equation}
(U_f(L))_{e\mu} = \frac{\alpha^\ast - \alpha}{2} \sin 2\theta + \beta
\cos 2\theta = -i \phi \frac{x_{21}}{L \tilde{\lambda}} \sin 2\theta
\sin (L \tilde{\lambda}).
\label{eq:emu}
\end{equation}
Taking the absolute value and squaring Eq.~(\ref{eq:emu}), we obtain
the transition probability $P_{e\mu}$ as
\begin{eqnarray}
P_{e\mu} &=& \vert \left(U_f(L)\right)_{e\mu} \vert^2 =
\frac{x_{21}^2}{(L \tilde{\lambda})^2} \sin^2 2\theta \sin^2 (L
\tilde{\lambda}) \nonumber\\
&=& \frac{x_{21}^2 \sin^2 2\theta}{L^2
\left[A^2/4 - x_{21} \left(A \cos 2\theta - x_{21}/L\right)/L \right]}
\sin^2 (L \tilde{\lambda}) \nonumber\\
&=& \frac{\sin^2 2\theta}{1 + AL \left[AL/(4
x_{21}) - \cos 2\theta\right]/x_{21}} \sin^2 (L \tilde{\lambda})
\nonumber\\
&=& \frac{\sin^2 2\theta}{\sin^2 2\theta + \left[ \cos 2\theta -
AL/(2 x_{21}) \right]^2} \sin^2 (L \tilde{\lambda}),
\end{eqnarray}
where $L \tilde{\lambda} = \sqrt{x_{21}^2 - AL x_{21} \left[ \cos
2\theta - AL/(4 x_{21}) \right]}$.
Introducing the definitions
\begin{equation}
\sin^2 2\theta^M \equiv \frac{\sin^2 2\theta}{\sin^2 2\theta + \left(
\cos 2\theta - A/E_{21}\right)^2}
\label{eq:thetaM}
\end{equation}
and
\begin{equation}
\tilde{x}_{21} \equiv \sqrt{x_{21}^2 - AL x_{21} \left( \cos
2\theta - \frac{AL}{4 x_{21}} \right)} = x_{21} \sqrt{\sin^2 2\theta +
\left( \cos 2\theta - \frac{A}{E_{21}} \right)^2},
\label{eq:tx21}
\end{equation}
we can write
\begin{equation}
P_{e\mu} = \sin^2 2\theta^M \sin^2 \tilde{x}_{21}.
\label{eq:Pemu}
\end{equation}
One could also immediately identify the definitions~(\ref{eq:thetaM})
and (\ref{eq:tx21}) from Eq.~(\ref{eq:emu}) by observing that the factor
$x_{21} \sin 2\theta/(L\tilde{\lambda})$ should correspond to $\sin
2\theta^M$ and that the argument $L \tilde{\lambda}$ in the factor $\sin (L
\tilde{\lambda})$ should correspond to $\tilde{x}_{21}$. This
identification corresponds to the one in Eq.~(\ref{eq:uum1}) above in
Sec.~\ref{sec:mma}. The factor $-i \phi$ is just a complex phase factor.

Setting
\begin{equation}
\cos 2\theta = \frac{A}{E_{21}}
\label{eq:MSW}
\end{equation}
in Eqs.~(\ref{eq:thetaM}) and (\ref{eq:Pemu}) gives
\begin{equation}
\sin^2 2\theta^M = 1 \quad \mbox{(maximal mixing)},
\end{equation}
\begin{equation}
P_{e\mu} = \sin^2 \sqrt{\left(x_{21} + \frac{AL}{2}\right)\left(x_{21}
- \frac{AL}{2}\right)} = \sin^2 \left( x_{21}
\sqrt{\left(1+\frac{A}{E_{21}}\right)\left(1-\frac{A}{E_{21}}\right)} \right),
\end{equation}
where Eq.~(\ref{eq:MSW}) is the so-called MSW resonance condition.

Inserting $A=0$ into Eq.~(\ref{eq:Pemu}) leads, of course, back to
$P_{e\mu} = \sin^2 2\theta \sin^2 x_{21}$, the old transition
probability formula for $\nu_e$-$\nu_\mu$ oscillations in vacuum.

\newpage

\end{document}